\documentclass[prl, twocolumn, 10pt, showpacs, aps, superscriptaddress]{revtex4-1}
\usepackage{amssymb, amsmath, hyperref, dsfont, color, graphicx, bm, upgreek}

\begin{document}

\title{Distinguishing Quantum and Classical Baths via Correlation Measurements}
\author{Thomas Fink}
\altaffiliation{Present address: Institute for Quantum Electronics, ETH Z\"urich, CH-8093 Z\"urich, Switzerland}
\affiliation{Department of Physics, RWTH Aachen University, D-52074 Aachen, Germany}
\affiliation{JARA - Fundamentals of Future Information Technology, D-52425 J\"ulich, Germany}
\author{Hendrik Bluhm}
\affiliation{Department of Physics, RWTH Aachen University, D-52074 Aachen, Germany}
\affiliation{JARA - Fundamentals of Future Information Technology, D-52425 J\"ulich, Germany}
\begin{abstract}
Investigations of quantum mechanical effects in macroscopic systems are of great interest to shed light onto the question where and how the transition to the classical world appears. It is also of practical relevance to determine if a bath dephasing a qubit can be described classically or requires a quantum mechanical treatment. We propose a measurement scheme to detect quantum backaction via correlation measurements to answer this question for a bath coupled to a single qubit. The presence of backaction leads to a dependence of correlations of subsequent initialization-evolution-readout cycles on how the qubit is manipulated in between. We compute the autocorrelation function for both an instructive spin model and the realistic case of an electron spin coupled to a bath of $10^6$ nuclear spins, as found in gated GaAs quantum dots, and show that backaction from the qubit onto the nuclear spin bath should be detectable even in such a large system.
\end{abstract}


\maketitle

Motivated by both fundamental questions and the promise of quantum computing, the decoherence of a qubit (or other quantum system) due to its interaction with a bath has received tremendous attention. One possibility to model this process is a fully quantum mechanical description, where a Hamiltonian is introduced for the complete system consisting of the qubit and its bath. While this approach is very general, the solution of the resulting many-body problem is usually challenging \cite{Stanek2013}. Experimentally, one can typically only control and measure the qubit, which makes it difficult to obtain information about the bath on a similar level of detail as contained in a quantum model. Hence, a much different approach has been widely used for practical purposes by describing the bath as a classical fluctuating field. The fluctuations of this field are often assumed to be a stationary random process and to exhibit Gaussian statistics, so that it is fully characterized by its noise spectral density. Based on this approach, a very powerful formalism connecting the loss of coherence to the noise spectrum via filter functions has been developed \cite{Cywinski2008}. On the one hand, it allows to predict the qubit decoherence due to the fluctuating bath, possibly under the application of pulse sequences. Much attention has also been paid to tailoring these sequences to optimally protect qubit coherence \cite{Viola1999,Uhrig2007,Viola1998,Khodjasteh2007}. On the other hand, analyzing the measured loss of coherence can be used to extract the underlying noise spectrum.

It is thus of great interest under what conditions the classical description is a good approximation to a full quantum model \cite{Stanek2013,Witzel2013}. Broadly speaking, the question is whether a bath behaves quantum mechanically or classically. Yet, fairly little work has been done to experimentally probe this question in a general manner. In this Letter, we propose a qualitative, experimental test to discriminate these regimes. Our method is based on measuring correlations of subsequent qubit readouts framing an intermediate interaction that causes backaction from the qubit onto the bath. Measurement of a backaction-induced change in the autocorrelation function then reveals whether a quantum mechanical treatment is required to describe the qubit-bath interactions. As a concrete and practically relevant example, we consider an electron spin qubit in a GaAs quantum dot, which is coupled to a bath of $10^4$ to $10^6$ nuclear spins of the host material \cite{Petta2005}, and show that the backaction of the qubit should be observable with established experimental methods on the time scale of the qubit's coherence time \cite{Bluhm2010}. The decoherence caused by this bath has received significant attention \cite{Chekhovich2013}. While quantum models have been considered in detail \cite{Witzel2006,Yao2006}, all experiments to date can be explained in a classical picture of the bath \cite{Petta2005,Koppens2006,Bluhm2010,Neder2011,Biercuk2011,Press2010}.

Given the mesoscopic nature of the nuclear spin bath, the question whether quantum effects can be observed is particularly pertinent. So far, there appears to be no fundamental bound on the validity of quantum mechanics for the macroscopic world other than the increase of decoherence. Nevertheless, it is a valid scientific question whether deviations can be observed as larger entities are considered \cite{Leggett2002}. Our proposed experiment would constitute such a test in the sense that a failure to observe the expected backaction, which is required for the uncertainty relation to hold, could point to a breakdown of quantum mechanics. For nitrogen-vacancy centers coupled to a bath consisting of only a few nuclear spins in diamond, this question has been addressed both theoretically and experimentally. The emergence of the quantum regime has been shown to appear when the electron spin-induced Knight shift is of comparable size to the externally applied magnetic field using a technique based on the particular energy level structure of this system \cite{Reinhard2012}, and by comparing dephasing between different states \cite{Zhao2011,Huang2011}. Our approach is more general as it makes fewer assumptions about the nature of the qubit and its coupling to the bath.

Note that the meaning of the term quantum bath is not unique, ranging from the dominance of quantum fluctuations \cite{Clerk2010} to the importance of internal dynamics \cite{Fedorov2011}. Here, we define a classical bath as one for which the dephasing of a single qubit can be described by a Hamiltonian of the form
\begin{equation} \label{eq:class}
\hat{H} = \hat{H}_Q + \bm{\beta}(t) \cdot \hat{\bm{\sigma}},
\end{equation}
where $\hat{H}_Q$ is a deterministic operator acting on the qubit only, $\hat{\bm{\sigma}} = (\hat{\sigma}^x,\hat{\sigma^y},\hat{\sigma}^z)$ are Pauli matrices and $\bm{\beta}(t)$ is the randomly fluctuating field caused by the bath. Being classical, this field is represented by real numbers rather than an operator. A quantum bath would be one where this description fails and one has to resort to the more general quantum model
\begin{equation} \label{eq:qm}
\hat{H} = \hat{H}_Q + \hat{H}_B + \hat{\bm{A}} \cdot \hat{\bm{\sigma}},
\end{equation}
where $\hat{H}_B$ describes the internal bath dynamics and $\hat{\bm{A}} \cdot \hat{\bm{\sigma}}$ denotes the quantum mechanical coupling between the qubit and the bath. In principle, one might distinguish these cases by testing a range of models. However, it will be difficult to rule out the existence of a suitable and possibly unknown classical model if only a quantum model is found to agree. Thus, we seek a qualitative, experimentally testable criterion.

We note that a key difference between the two scenarios is that in Eq.~\ref{eq:class}, the dynamics of the bath are independent of the state of the qubit whereas in Eq.~\ref{eq:qm}, the coupling term implies that the evolution of the qubit state also affects the bath dynamics. In other words, the bath experiences a backaction from the qubit. We thus use the observation of backaction as the operational criterion for a bath to be quantum. As a further constraint, one may demand that this backaction should be detectable on the time scale of the qubit's coherence time. For example, repeatedly transferring energy from the qubit to the bath will eventually heat any finite size bath, but this form of backaction would hardly be considered a quantum effect.

As a test for a detectable backaction, we propose to measure correlations of repeated (near single-shot) measurements of the qubit, which we recently introduced as an alternative way to characterize the noise spectrum of a bath assumed to be classical \cite{Fink2013}. The scheme consists of two free evolution periods in each of which the qubit is initialized into a $\hat{\sigma}^x$ eigenstate, evolves for a time $t_M$ under the influence of the coupling Hamiltonian $\hat{H}_M$ and is then measured projectively in the $\hat{\sigma}^x$ basis. For the sake of simplicity, we assume initialization and measurement to be negligibly short compared to the evolution times of interest.

\begin{figure}[t]
\includegraphics{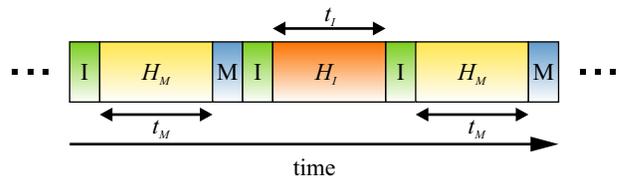}
\caption{Measurement cycle: Qubit initialization into a superposition state (I), evolution under the respective coupling Hamiltonian $\hat{H}_{M,I}$ for time $t_{M,I}$, and projective measurement (M). The correlation of the two qubit readouts after the commuting evolutions $\hat{H}_M$ will then measure the effect of the possibly noncommuting intermediate qubit-bath coupling operator $\hat{H}_I$.}\label{fig:protocol}
\end{figure}

Correlations of subsequent projective readouts measure the intermediate bath dynamics via correlations of the phases accumulated by the qubit. To make this protocol sensitive to the backaction from the qubit onto the bath, we incorporate an intermediate qubit-bath-interaction described by $\hat{H}_I$ after the first projective measurement. This sequence is depicted in Fig.~\ref{fig:protocol}.
Any effect of the intermediate interaction of the qubit with the bath on the correlations between the two measurements is an unambiguous proof of backaction (provided, of course, there is no spurious effect of the qubit control field coupling directly to the bath). This picture can be motivated as follows: For simplicity, assume no internal dynamics of the bath. Then, each interaction-readout sequence (including the intermediate interaction, independent of whether the qubit is actually read out or not) can be considered as a measurement of some bath operator (with a binary outcome). Without the intermediate measurements, we are facing two consecutive projective measurements of the same observable, which should be perfectly correlated. However, if the intermediate measurement $\hat{H}_I$ does not commute with the outer ones $\hat{H}_M$, the laws of quantum mechanics require that such a measurement does have a backaction, which ensures that the uncertainty relation for non-commuting observables holds. Thus, the correlation would be reduced since the backaction would drive the bath out of the eigenspace into which it is projected in the first measurement.

To analyze the protocol, we compute the correlation between the measurements. While it is generally independent of the form of the coupling Hamiltonian, we restrict our analysis to the case where $\hat{H} \propto \hat{\sigma}^z$ and $A^x = A^y = 0$ for simplicity. We first consider a pure initial bath state $|J\rangle$ to later generalize to mixed states by averaging. After initializing the qubit into a superposition $|\pm\rangle = \frac{1}{\sqrt{2}} (|0\rangle \pm |1\rangle)$, where $|0\rangle$ and $|1\rangle$ are the $\hat{\sigma}^z$ eigenstates, the coupling operator in Eq.~\ref{eq:qm} starts to entangle the qubit and bath subspace leading to a state
\begin{equation} \label{eq:bathstate}
|\psi(t)\rangle = \frac{1}{\sqrt{2}} \left( |0\rangle \otimes \hat{U}_0(t) |J\rangle + |1\rangle \otimes \hat{U}_1(t) |J\rangle \right),
\end{equation}
where $\hat{U}_i(t) = \hat{\mathcal{T}} \exp \left(-i \int_0^{t} dt^\prime \hat{H}_i(t^\prime) /\hbar \right)$ denotes the operators mediating entanglement between the bath state and the two qubit states $|0\rangle$ and $|1\rangle$. $\hat{\mathcal{T}}$ is the time-ordering operator. A subsequent measurement of the qubit projects the  bath onto a quantum mechanical superposition state
\begin{equation} \label{eq:entanglement}
|J(t)\rangle = \frac{1}{\sqrt{2}} \left( \hat{U}_0(t) \pm \hat{U}_1(t) \right) |J\rangle,
\end{equation}
where the sign is determined by the outcome of the qubit measurement in the superposition basis. Note that terms of the form $\langle J| \hat{U}_0^\dagger(t) \hat{U}_1(t) |J\rangle$ have been neglected in the normalization of Eq.~\ref{eq:entanglement} as they are small for time scales much larger than the free induction decay (FID). We find that the autocorrelation $C = \langle M_1 \cdot M_2 \rangle$ between two measurement values $M_{1,2}$ for the sequence depicted in Fig.~\ref{fig:protocol} is given by
\begin{eqnarray} \nonumber \label{eq:correlation}
&C& = \frac{1}{2} \langle J | \hat{U}_0^\dagger(t_M) \hat{U}_I^\dagger(t_I) \hat{U}_0^\dagger(t_M) \hat{U}_1(t_M) \hat{U}_I(t_I) \hat{U}_1(t_M) | J \rangle \\
&+& \frac{1}{2} \langle J | \hat{U}_1^\dagger(t_M) \hat{U}_I^\dagger(t_I) \hat{U}_0^\dagger(t_M) \hat{U}_1(t_M) \hat{U}_I(t_I) \hat{U}_0(t_M) | J \rangle.
\end{eqnarray}
The relation between measurement correlations and the different evolution operators is the main general result of this paper. It is now apparent that, if $\hat{H}_{0,1}$ and $\hat{H}_I$ commute, $\hat{U}_I$ can be eliminated so that the value of $C$ is independent of what happens during the intermediate evolution and no backaction is detected. Otherwise, the value of $C$ in general depends on how the qubit is manipulated during $\hat{U}_I$, which is an unambiguous evidence that its backaction affects the bath. One can further show that if measurement and intermediate evolution time are negligibly short so that $\hat{U}_I = \mathds{1}$, this result corresponds to the Hahn echo amplitude plus some rapidly decaying terms. We had previously obtained this result for classical Gaussian noise \cite{Fink2013}.

We now turn to the central spin model as a concrete example and will consider a prominent realization consisting of two coupled electrically confined electron spins embedded in a GaAs host lattice. This so-called singlet-triplet qubit \cite{Petta2005} exhibits dephasing times up to 200\,$\upmu$s \cite{Bluhm2010} and allows fast single-shot readout within 100\,ns \cite{Barthel2010}, thus making it a good candidate for applying our scheme. It is instructive to first consider a situation where a central electron spin is coupled to a bath of $N$ nuclei of the same species (e.g. one Gallium isotope) with spin $\frac{1}{2}$. We assume that the nuclear spins are in an infinite temperature mixed state. For external magnetic fields much larger than the hyperfine coupling, spin flips between the electron and nuclei are negligible and we can replace the coupling operator in Eq.~\ref{eq:qm} by $\hbar \sum_i^N A_i \hat{S}^z \hat{J}_i^z$. Here, $\hat{S}^z$ ($\hat{J}_i^z$) is the electron spin ($i$th nuclear spin) operator and $A_i = \mathcal{A} \times | \psi(\mathbf{r}_i) |^2$ describes the coupling strength of the $i$th nuclear spin at position $\mathbf{r}_i$ weighted by the electron wave function $\psi(\mathbf{r})$ \cite{Cywinski2009}. A direct approach to realize a noncommuting evolution during the intermediate period is the rotation of the electron spin quantization axis, e.g. by flipping the external magnetic field into the $x$-direction. This effectively couples the electron spin to the $\hat{J}_i^x$ component of each nuclear spin such that $\hat{U}_{0,1} = \otimes_i^N \exp \left( \mp \frac{1}{2} A_i \hat{J}_i^z t_M \right)$ and $\hat{U}_I = \otimes_i^N \exp \left( - \frac{1}{2} A_i \hat{J}_i^x t_I \right)$ do not commute. The derivation of the correlation function for this case and the straightforward extension to the double dot configuration in the singlet-triplet qubit is presented in the supplemental material \footnote{See Supplemental Material at [URL] for a detailed derivation of the autocorrelation function for the switched unixaxial coupling and isotropic coupling with dynamically decoupling pulses scheme.}.

\begin{figure}[t]
\includegraphics{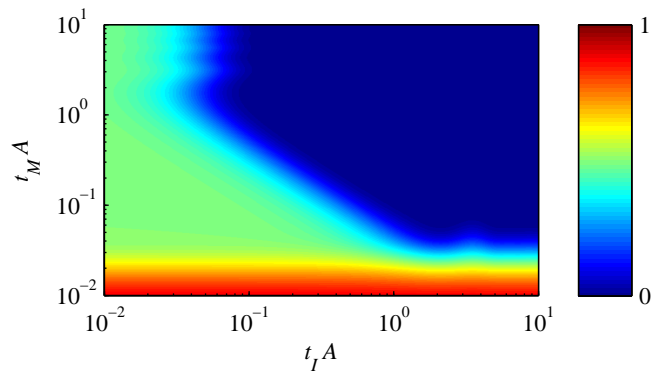}
\caption{Autocorrelation function for the switched uniaxial coupling scheme applied to the singlet-triplet qubit with $10^3$ nuclear spins of the same species in each dot. The correlation between the two measurements decays as a function of the intermediate evolution time when $t_M A \lesssim 1$.} \label{fig:toymodel}
\end{figure}
The result is illustrated in Fig.~\ref{fig:toymodel} for symmetric dots with each $10^3$ nuclei of the same species. In the regime where the outer evolution times are small compared to the inverse hyperfine interaction, the correlation between the two qubit measurements decreases as a function of the intermediate evolution time, which shows that backaction from the qubit onto the noise bath can indeed be detected. In this regime, the intermediate interaction rotates the nuclear spins about the $x$-axis, and thus erases the projection effect of the first measurement along the $z$-axis. For homogeneous coupling, one would expect revivals when the electron-spin induced Knight shift causes a full rotation of the nuclei. However, this effect is completely washed out due to the inhomogeneous coupling \footnotemark[\value{footnote}] and only leads to the small beating features at $t_M A = t_M \mathcal{A}/N \sim 1$ and $t_I A \sim 1$. Moreover, in the experimentally relevant regime $t_M A \lesssim 1$, the outer evolution times set the time scale on which the correlation decays which allows for tuning of the backaction-induced decay to time scales where other decay mechanisms, such as inhomogeneous broadening, are still negligible. For $t_M A \ll 1$, the two measurements detect the initialized states leading to a perfect correlation regardless of the intermediate interaction if those states are identical. Changing the number of nuclear spins exhibits a qualitatively similar behavior with a shifted decay of correlation. This can be seen from the relation $t \times \tau \propto N^{3/2}$, which holds at the point where $C = \frac{1}{2}e^{-1}$, i.e. the diagonal transition between high and low correlations in Fig.~\ref{fig:toymodel}.

While the rapid change of the quantization axis considered above provides an illustrative example for generating two noncommuting evolutions, it is nontrivial to realize such a fast reorientation of the magnetic field direction in practice. Therefore, we now consider a more accessible approach based on applying different pulse sequences to the electron. For a FID experiment, which consists of the initialization of the qubit into a $\hat{\sigma}^x$ eigenstate, a free evolution for a given time and a subsequent projective measurement, the secular terms of the hyperfine coupling between the electron and each nucleus are dominant. This makes the final qubit measurement predominantly sensitive to the Overhauser field created by the nuclei along the qubit quantization axis
\begin{equation}
g \mu_B B_{nuc}^z = \hbar \sum_i^N A_i J_i^z,
\end{equation}
which is assumed to be constant on typical qubit evolution time scales. During a spin echo (SE) experiment, a $\pi$-pulse applied to the qubit after half of the evolution time inverts the direction of the hyperfine coupling, effectively reversing any constant interaction during the second half of the evolution time. In contrast, second order terms make the SE decay sensitive to the time-dependent transverse Overhauser field components $B_{nuc}^\bot(t)$ that oscillate as a result of the relative Larmor precession of the different isotopes in the system \cite{Bluhm2010}. Since $\hat{J}_i^z$ and $\hat{J}_i^\pm = \hat{J}_i^x \pm i \hat{J}_i^y$ do not commute, implementing the two outer evolutions in Fig.~\ref{fig:protocol} as a SE and the intermediate evolution as a FID effectively creates the required noncommuting evolutions. Differences observed in comparing this SE-FID-SE scheme to a SE-SE-SE scheme with the same durations will then allow to  directly identify a potential backaction. We make use of a semiclassical approach \cite{Neder2011} to calculate the autocorrelation function for our measurement scheme \footnotemark[\value{footnote}], and again focus on the singlet-triplet qubit where similar features as we predict have already been observed in standard SE experiments \cite{Bluhm2010}. However, our results are equally valid for single spins.

\begin{figure}[t]
\includegraphics{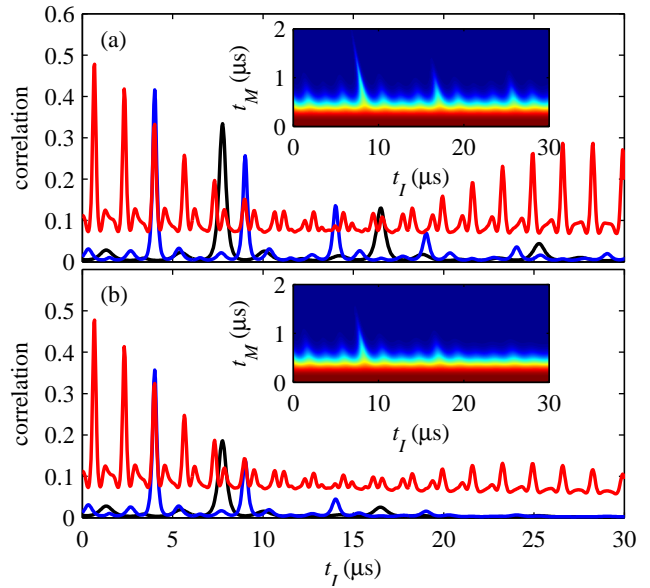}
\caption{(a) Autocorrelation function with intermediate SE for external magnetic fields of 40, 70, and 210\,mT (black, blue, and red curve, respectively) and $\sqrt{\langle \delta B \rangle^2} = 0.2$\,mT. The outer evolution time is $t_M=$1\,$\upmu$s and each electron is coupled to $10^6$ nuclei consisting of $^{69}$Ga, $^{71}$Ga, and $^{75}$As. Inset: Correlation as colorscale as a function of outer and intermediate evolution time at external magnetic field of 40\,mT. (b) Same as (a) with intermediate FID. The suppression of the revivals due to the intermediate backaction are clearly visible.} \label{fig:sefid}
\end{figure}
The results are depicted in Fig.~\ref{fig:sefid} and show revivals whenever all three species have the same relative orientation during both measurements, i.e. a multiple of the relative precession periods has elapsed. This condition can be met due to the fortuitous spacing of the nuclear Larmor frequencies in GaAs. To phenomenologically account for nuclear dephasing we have included Larmor frequency deviations due to local Gaussian magnetic field deviations $\delta B$ leading to an overall decay of the correlation function \cite{Neder2011}. $\delta B$ is chosen slightly smaller than extracted from earlier experiments \cite{Bluhm2010}, expecting that it can be reduced by optimizing the direction of the external field. Remarkably, turning on the backaction by introducing the intermediate FID leads to a clear suppression of the correlation revivals which stems from the backaction of the electron as compared to the intermediate SE. This is due to the Knight shift experienced by the nuclei during the FID which changes the relative phase of the Larmor precessions. For an SE, however, this phase shift is echoed away. Careful tuning of the external magnetic field will then allow to observe this backaction-induced revival suppression on time scales where decoherence due to dipolar interactions, which occurs on a timescale of about 30\,$\upmu$s \cite{Bluhm2010}, does not render them invisible. In principle, inverting the sequence by exchanging FID and SE equally fulfills the noncommutation condition. However, the second measurement is then sensitive to the backaction-induced change of the longitudinal component of the Overhauser field which is for the sake of simplicity not accounted for in our model.

In conclusion, we have presented a measurement scheme to qualitatively determine whether a bath behaves classical or quantum mechanical via readout of a qubit interacting with it. One may ask if backaction from the qubit onto a coupled bath can also be described with purely classical models. In principle, this is indeed the case (e.g., by treating each nuclear spin as a vector with a rotation axis determined by the qubit state). However, the projective measurement of the qubit in the superposition basis entangles it with the nuclear spin bath, which is a purely quantum mechanical phenomenon. We note that any failure to observe the expected backaction of a noncommuting intermediate measurement would constitute a violation of the very principles of quantum mechanics that enforce the uncertainty relation. In this sense, the experiment we propose can be seen as a test of the validity of quantum mechanics for fairly macroscopic entities, such as an ensemble of $10^6$ nuclear spins.

We speculate that the observation of backaction from a qubit on the scale of its coherence time also implies that the qubit-bath system can be put into an entangled state even though the bath starts in an infinite temperature state. One can show that for an infinite temperature bath and a Hamiltonian given by Eq.~\ref{eq:qm} but only involving terms proportional to $\hat{\sigma}^z$, no entanglement can be formed with a single interaction period. However, the first projective measurement would leave the bath in a state that is not maximally mixed and might thus allow the second interaction period to create entanglement.

\acknowledgments{This work was supported by the Alfried Krupp von Bohlen und Halbach Foundation. We would like to thank Lorenza Viola for useful discussions.}

\bibliography{backaction}

\begin{thebibliography}{26}%
\makeatletter
\providecommand \@ifxundefined [1]{%
 \@ifx{#1\undefined}
}%
\providecommand \@ifnum [1]{%
 \ifnum #1\expandafter \@firstoftwo
 \else \expandafter \@secondoftwo
 \fi
}%
\providecommand \@ifx [1]{%
 \ifx #1\expandafter \@firstoftwo
 \else \expandafter \@secondoftwo
 \fi
}%
\providecommand \natexlab [1]{#1}%
\providecommand \enquote  [1]{``#1''}%
\providecommand \bibnamefont  [1]{#1}%
\providecommand \bibfnamefont [1]{#1}%
\providecommand \citenamefont [1]{#1}%
\providecommand \href@noop [0]{\@secondoftwo}%
\providecommand \href [0]{\begingroup \@sanitize@url \@href}%
\providecommand \@href[1]{\@@startlink{#1}\@@href}%
\providecommand \@@href[1]{\endgroup#1\@@endlink}%
\providecommand \@sanitize@url [0]{\catcode `\\12\catcode `\$12\catcode
  `\&12\catcode `\#12\catcode `\^12\catcode `\_12\catcode `\%12\relax}%
\providecommand \@@startlink[1]{}%
\providecommand \@@endlink[0]{}%
\providecommand \url  [0]{\begingroup\@sanitize@url \@url }%
\providecommand \@url [1]{\endgroup\@href {#1}{\urlprefix }}%
\providecommand \urlprefix  [0]{URL }%
\providecommand \Eprint [0]{\href }%
\providecommand \doibase [0]{http://dx.doi.org/}%
\providecommand \selectlanguage [0]{\@gobble}%
\providecommand \bibinfo  [0]{\@secondoftwo}%
\providecommand \bibfield  [0]{\@secondoftwo}%
\providecommand \translation [1]{[#1]}%
\providecommand \BibitemOpen [0]{}%
\providecommand \bibitemStop [0]{}%
\providecommand \bibitemNoStop [0]{.\EOS\space}%
\providecommand \EOS [0]{\spacefactor3000\relax}%
\providecommand \BibitemShut  [1]{\csname bibitem#1\endcsname}%
\let\auto@bib@innerbib\@empty
\bibitem [{\citenamefont {Stanek}\ \emph {et~al.}(2013)\citenamefont {Stanek},
  \citenamefont {Raas},\ and\ \citenamefont {Uhrig}}]{Stanek2013}%
  \BibitemOpen
  \bibfield  {author} {\bibinfo {author} {\bibfnamefont {D.}~\bibnamefont
  {Stanek}}, \bibinfo {author} {\bibfnamefont {C.}~\bibnamefont {Raas}}, \ and\
  \bibinfo {author} {\bibfnamefont {G.~S.}\ \bibnamefont {Uhrig}},\ }\href
  {\doibase 10.1103/PhysRevB.88.155305} {\bibfield  {journal} {\bibinfo
  {journal} {Phys. Rev. B}\ }\textbf {\bibinfo {volume} {88}},\ \bibinfo
  {pages} {155305} (\bibinfo {year} {2013})}\BibitemShut {NoStop}%
\bibitem [{\citenamefont {Cywi\ifmmode~\acute{n}\else \'{n}\fi{}ski}\ \emph
  {et~al.}(2008)\citenamefont {Cywi\ifmmode~\acute{n}\else \'{n}\fi{}ski},
  \citenamefont {Lutchyn}, \citenamefont {Nave},\ and\ \citenamefont
  {Das~Sarma}}]{Cywinski2008}%
  \BibitemOpen
  \bibfield  {author} {\bibinfo {author} {\bibfnamefont {L.}~\bibnamefont
  {Cywi\ifmmode~\acute{n}\else \'{n}\fi{}ski}}, \bibinfo {author}
  {\bibfnamefont {R.~M.}\ \bibnamefont {Lutchyn}}, \bibinfo {author}
  {\bibfnamefont {C.~P.}\ \bibnamefont {Nave}}, \ and\ \bibinfo {author}
  {\bibfnamefont {S.}~\bibnamefont {Das~Sarma}},\ }\href {\doibase
  10.1103/PhysRevB.77.174509} {\bibfield  {journal} {\bibinfo  {journal} {Phys.
  Rev. B}\ }\textbf {\bibinfo {volume} {77}},\ \bibinfo {pages} {174509}
  (\bibinfo {year} {2008})}\BibitemShut {NoStop}%
\bibitem [{\citenamefont {Viola}\ \emph {et~al.}(1999)\citenamefont {Viola},
  \citenamefont {Knill},\ and\ \citenamefont {Lloyd}}]{Viola1999}%
  \BibitemOpen
  \bibfield  {author} {\bibinfo {author} {\bibfnamefont {L.}~\bibnamefont
  {Viola}}, \bibinfo {author} {\bibfnamefont {E.}~\bibnamefont {Knill}}, \ and\
  \bibinfo {author} {\bibfnamefont {S.}~\bibnamefont {Lloyd}},\ }\href
  {\doibase 10.1103/PhysRevLett.82.2417} {\bibfield  {journal} {\bibinfo
  {journal} {Phys. Rev. Lett.}\ }\textbf {\bibinfo {volume} {82}},\ \bibinfo
  {pages} {2417} (\bibinfo {year} {1999})}\BibitemShut {NoStop}%
\bibitem [{\citenamefont {Uhrig}(2007)}]{Uhrig2007}%
  \BibitemOpen
  \bibfield  {author} {\bibinfo {author} {\bibfnamefont {G.~S.}\ \bibnamefont
  {Uhrig}},\ }\href {\doibase 10.1103/PhysRevLett.98.100504} {\bibfield
  {journal} {\bibinfo  {journal} {Phys. Rev. Lett.}\ }\textbf {\bibinfo
  {volume} {98}},\ \bibinfo {pages} {100504} (\bibinfo {year}
  {2007})}\BibitemShut {NoStop}%
\bibitem [{\citenamefont {Viola}\ and\ \citenamefont
  {Lloyd}(1998)}]{Viola1998}%
  \BibitemOpen
  \bibfield  {author} {\bibinfo {author} {\bibfnamefont {L.}~\bibnamefont
  {Viola}}\ and\ \bibinfo {author} {\bibfnamefont {S.}~\bibnamefont {Lloyd}},\
  }\href {\doibase 10.1103/PhysRevA.58.2733} {\bibfield  {journal} {\bibinfo
  {journal} {Phys. Rev. A}\ }\textbf {\bibinfo {volume} {58}},\ \bibinfo
  {pages} {2733} (\bibinfo {year} {1998})}\BibitemShut {NoStop}%
\bibitem [{\citenamefont {Khodjasteh}\ and\ \citenamefont
  {Lidar}(2007)}]{Khodjasteh2007}%
  \BibitemOpen
  \bibfield  {author} {\bibinfo {author} {\bibfnamefont {K.}~\bibnamefont
  {Khodjasteh}}\ and\ \bibinfo {author} {\bibfnamefont {D.~A.}\ \bibnamefont
  {Lidar}},\ }\href {\doibase 10.1103/PhysRevA.75.062310} {\bibfield  {journal}
  {\bibinfo  {journal} {Phys. Rev. A}\ }\textbf {\bibinfo {volume} {75}},\
  \bibinfo {pages} {062310} (\bibinfo {year} {2007})}\BibitemShut {NoStop}%
\bibitem [{\citenamefont {Witzel}\ \emph {et~al.}(2013)\citenamefont {Witzel},
  \citenamefont {Young},\ and\ \citenamefont {Das~Sarma}}]{Witzel2013}%
  \BibitemOpen
  \bibfield  {author} {\bibinfo {author} {\bibfnamefont {W.~M.}\ \bibnamefont
  {Witzel}}, \bibinfo {author} {\bibfnamefont {K.}~\bibnamefont {Young}}, \
  and\ \bibinfo {author} {\bibfnamefont {S.}~\bibnamefont {Das~Sarma}},\
  }\href@noop {} {\enquote {\bibinfo {title} {Converting a real quantum bath to
  an effective classical noise},}\ } (\bibinfo {year} {2013}),\ \Eprint
  {http://arxiv.org/abs/1307.2597} {arXiv:1307.2597} \BibitemShut {NoStop}%
\bibitem [{\citenamefont {Petta}\ \emph {et~al.}(2005)\citenamefont {Petta},
  \citenamefont {Johnson}, \citenamefont {Taylor}, \citenamefont {Laird},
  \citenamefont {Yacoby}, \citenamefont {Lukin}, \citenamefont {Marcus},
  \citenamefont {Hanson},\ and\ \citenamefont {Gossard}}]{Petta2005}%
  \BibitemOpen
  \bibfield  {author} {\bibinfo {author} {\bibfnamefont {J.~R.}\ \bibnamefont
  {Petta}}, \bibinfo {author} {\bibfnamefont {A.~C.}\ \bibnamefont {Johnson}},
  \bibinfo {author} {\bibfnamefont {J.~M.}\ \bibnamefont {Taylor}}, \bibinfo
  {author} {\bibfnamefont {E.~A.}\ \bibnamefont {Laird}}, \bibinfo {author}
  {\bibfnamefont {A.}~\bibnamefont {Yacoby}}, \bibinfo {author} {\bibfnamefont
  {M.~D.}\ \bibnamefont {Lukin}}, \bibinfo {author} {\bibfnamefont {C.~M.}\
  \bibnamefont {Marcus}}, \bibinfo {author} {\bibfnamefont {M.~P.}\
  \bibnamefont {Hanson}}, \ and\ \bibinfo {author} {\bibfnamefont {A.~C.}\
  \bibnamefont {Gossard}},\ }\href {\doibase 10.1126/science.1116955}
  {\bibfield  {journal} {\bibinfo  {journal} {Science}\ }\textbf {\bibinfo
  {volume} {309}},\ \bibinfo {pages} {2180} (\bibinfo {year}
  {2005})}\BibitemShut {NoStop}%
\bibitem [{\citenamefont {Bluhm}\ \emph {et~al.}(2010)\citenamefont {Bluhm},
  \citenamefont {Foletti}, \citenamefont {Neder}, \citenamefont {Rudner},
  \citenamefont {Mahalu}, \citenamefont {Umansky},\ and\ \citenamefont
  {Yacoby}}]{Bluhm2010}%
  \BibitemOpen
  \bibfield  {author} {\bibinfo {author} {\bibfnamefont {H.}~\bibnamefont
  {Bluhm}}, \bibinfo {author} {\bibfnamefont {S.}~\bibnamefont {Foletti}},
  \bibinfo {author} {\bibfnamefont {I.}~\bibnamefont {Neder}}, \bibinfo
  {author} {\bibfnamefont {M.}~\bibnamefont {Rudner}}, \bibinfo {author}
  {\bibfnamefont {D.}~\bibnamefont {Mahalu}}, \bibinfo {author} {\bibfnamefont
  {V.}~\bibnamefont {Umansky}}, \ and\ \bibinfo {author} {\bibfnamefont
  {A.}~\bibnamefont {Yacoby}},\ }\href {\doibase 10.1038/nphys1856} {\bibfield
  {journal} {\bibinfo  {journal} {Nature Physics}\ }\textbf {\bibinfo {volume}
  {7}},\ \bibinfo {pages} {109} (\bibinfo {year} {2010})}\BibitemShut {NoStop}%
\bibitem [{\citenamefont {Chekhovich}\ \emph {et~al.}(2013)\citenamefont
  {Chekhovich}, \citenamefont {Makhonin}, \citenamefont {Tartakovskii},
  \citenamefont {Yacoby}, \citenamefont {Bluhm}, \citenamefont {Nowack},\ and\
  \citenamefont {Vandersypen}}]{Chekhovich2013}%
  \BibitemOpen
  \bibfield  {author} {\bibinfo {author} {\bibfnamefont {E.~A.}\ \bibnamefont
  {Chekhovich}}, \bibinfo {author} {\bibfnamefont {M.~N.}\ \bibnamefont
  {Makhonin}}, \bibinfo {author} {\bibfnamefont {A.~I.}\ \bibnamefont
  {Tartakovskii}}, \bibinfo {author} {\bibfnamefont {A.}~\bibnamefont
  {Yacoby}}, \bibinfo {author} {\bibfnamefont {H.}~\bibnamefont {Bluhm}},
  \bibinfo {author} {\bibfnamefont {K.~C.}\ \bibnamefont {Nowack}}, \ and\
  \bibinfo {author} {\bibfnamefont {L.~M.~K.}\ \bibnamefont {Vandersypen}},\
  }\href {\doibase 10.1038/nmat3652} {\bibfield  {journal} {\bibinfo  {journal}
  {Nature Materials}\ }\textbf {\bibinfo {volume} {12}},\ \bibinfo {pages}
  {494} (\bibinfo {year} {2013})}\BibitemShut {NoStop}%
\bibitem [{\citenamefont {Witzel}\ and\ \citenamefont
  {Das~Sarma}(2006)}]{Witzel2006}%
  \BibitemOpen
  \bibfield  {author} {\bibinfo {author} {\bibfnamefont {W.~M.}\ \bibnamefont
  {Witzel}}\ and\ \bibinfo {author} {\bibfnamefont {S.}~\bibnamefont
  {Das~Sarma}},\ }\href {\doibase 10.1103/PhysRevB.74.035322} {\bibfield
  {journal} {\bibinfo  {journal} {Phys. Rev. B}\ }\textbf {\bibinfo {volume}
  {74}},\ \bibinfo {pages} {035322} (\bibinfo {year} {2006})}\BibitemShut
  {NoStop}%
\bibitem [{\citenamefont {Yao}\ \emph {et~al.}(2006)\citenamefont {Yao},
  \citenamefont {Liu},\ and\ \citenamefont {Sham}}]{Yao2006}%
  \BibitemOpen
  \bibfield  {author} {\bibinfo {author} {\bibfnamefont {W.}~\bibnamefont
  {Yao}}, \bibinfo {author} {\bibfnamefont {R.-B.}\ \bibnamefont {Liu}}, \ and\
  \bibinfo {author} {\bibfnamefont {L.~J.}\ \bibnamefont {Sham}},\ }\href
  {\doibase 10.1103/PhysRevB.74.195301} {\bibfield  {journal} {\bibinfo
  {journal} {Phys. Rev. B}\ }\textbf {\bibinfo {volume} {74}},\ \bibinfo
  {pages} {195301} (\bibinfo {year} {2006})}\BibitemShut {NoStop}%
\bibitem [{\citenamefont {Koppens}\ \emph {et~al.}(2006)\citenamefont
  {Koppens}, \citenamefont {Buizert}, \citenamefont {Tielrooij}, \citenamefont
  {Vink}, \citenamefont {Nowack}, \citenamefont {Meunier}, \citenamefont
  {Kouwenhoven},\ and\ \citenamefont {Vandersypen}}]{Koppens2006}%
  \BibitemOpen
  \bibfield  {author} {\bibinfo {author} {\bibfnamefont {F.~H.~L.}\
  \bibnamefont {Koppens}}, \bibinfo {author} {\bibfnamefont {C.}~\bibnamefont
  {Buizert}}, \bibinfo {author} {\bibfnamefont {K.~J.}\ \bibnamefont
  {Tielrooij}}, \bibinfo {author} {\bibfnamefont {I.~T.}\ \bibnamefont {Vink}},
  \bibinfo {author} {\bibfnamefont {K.~C.}\ \bibnamefont {Nowack}}, \bibinfo
  {author} {\bibfnamefont {T.}~\bibnamefont {Meunier}}, \bibinfo {author}
  {\bibfnamefont {L.~P.}\ \bibnamefont {Kouwenhoven}}, \ and\ \bibinfo {author}
  {\bibfnamefont {L.~M.~K.}\ \bibnamefont {Vandersypen}},\ }\href {\doibase
  10.1038/nature05065} {\bibfield  {journal} {\bibinfo  {journal} {Nature}\
  }\textbf {\bibinfo {volume} {442}},\ \bibinfo {pages} {766} (\bibinfo {year}
  {2006})}\BibitemShut {NoStop}%
\bibitem [{\citenamefont {Neder}\ \emph {et~al.}(2011)\citenamefont {Neder},
  \citenamefont {Rudner}, \citenamefont {Bluhm}, \citenamefont {Foletti},
  \citenamefont {Halperin},\ and\ \citenamefont {Yacoby}}]{Neder2011}%
  \BibitemOpen
  \bibfield  {author} {\bibinfo {author} {\bibfnamefont {I.}~\bibnamefont
  {Neder}}, \bibinfo {author} {\bibfnamefont {M.~S.}\ \bibnamefont {Rudner}},
  \bibinfo {author} {\bibfnamefont {H.}~\bibnamefont {Bluhm}}, \bibinfo
  {author} {\bibfnamefont {S.}~\bibnamefont {Foletti}}, \bibinfo {author}
  {\bibfnamefont {B.~I.}\ \bibnamefont {Halperin}}, \ and\ \bibinfo {author}
  {\bibfnamefont {A.}~\bibnamefont {Yacoby}},\ }\href {\doibase
  10.1103/PhysRevB.84.035441} {\bibfield  {journal} {\bibinfo  {journal} {Phys.
  Rev. B}\ }\textbf {\bibinfo {volume} {84}},\ \bibinfo {pages} {035441}
  (\bibinfo {year} {2011})}\BibitemShut {NoStop}%
\bibitem [{\citenamefont {Biercuk}\ and\ \citenamefont
  {Bluhm}(2011)}]{Biercuk2011}%
  \BibitemOpen
  \bibfield  {author} {\bibinfo {author} {\bibfnamefont {M.~J.}\ \bibnamefont
  {Biercuk}}\ and\ \bibinfo {author} {\bibfnamefont {H.}~\bibnamefont
  {Bluhm}},\ }\href {\doibase 10.1103/PhysRevB.83.235316} {\bibfield  {journal}
  {\bibinfo  {journal} {Phys. Rev. B}\ }\textbf {\bibinfo {volume} {83}},\
  \bibinfo {pages} {235316} (\bibinfo {year} {2011})}\BibitemShut {NoStop}%
\bibitem [{\citenamefont {Press}\ \emph {et~al.}(2010)\citenamefont {Press},
  \citenamefont {De~Greve}, \citenamefont {McMahon}, \citenamefont {Ladd},
  \citenamefont {Friess}, \citenamefont {Schneider}, \citenamefont {Kamp},
  \citenamefont {Hoefling}, \citenamefont {Forchel},\ and\ \citenamefont
  {Yamamoto}}]{Press2010}%
  \BibitemOpen
  \bibfield  {author} {\bibinfo {author} {\bibfnamefont {D.}~\bibnamefont
  {Press}}, \bibinfo {author} {\bibfnamefont {K.}~\bibnamefont {De~Greve}},
  \bibinfo {author} {\bibfnamefont {P.~L.}\ \bibnamefont {McMahon}}, \bibinfo
  {author} {\bibfnamefont {T.~D.}\ \bibnamefont {Ladd}}, \bibinfo {author}
  {\bibfnamefont {B.}~\bibnamefont {Friess}}, \bibinfo {author} {\bibfnamefont
  {C.}~\bibnamefont {Schneider}}, \bibinfo {author} {\bibfnamefont
  {M.}~\bibnamefont {Kamp}}, \bibinfo {author} {\bibfnamefont {S.}~\bibnamefont
  {Hoefling}}, \bibinfo {author} {\bibfnamefont {A.}~\bibnamefont {Forchel}}, \
  and\ \bibinfo {author} {\bibfnamefont {Y.}~\bibnamefont {Yamamoto}},\ }\href
  {\doibase 10.1038/NPHOTON.2010.83} {\bibfield  {journal} {\bibinfo  {journal}
  {Nature Photonics}\ }\textbf {\bibinfo {volume} {4}},\ \bibinfo {pages} {367}
  (\bibinfo {year} {2010})}\BibitemShut {NoStop}%
\bibitem [{\citenamefont {Leggett}(2002)}]{Leggett2002}%
  \BibitemOpen
  \bibfield  {author} {\bibinfo {author} {\bibfnamefont {A.~J.}\ \bibnamefont
  {Leggett}},\ }\href {http://stacks.iop.org/0953-8984/14/i=15/a=201}
  {\bibfield  {journal} {\bibinfo  {journal} {Journal of Physics: Condensed
  Matter}\ }\textbf {\bibinfo {volume} {14}},\ \bibinfo {pages} {R415}
  (\bibinfo {year} {2002})}\BibitemShut {NoStop}%
\bibitem [{\citenamefont {Reinhard}\ \emph {et~al.}(2012)\citenamefont
  {Reinhard}, \citenamefont {Shi}, \citenamefont {Zhao}, \citenamefont {Rempp},
  \citenamefont {Naydenov}, \citenamefont {Meijer}, \citenamefont {Hall},
  \citenamefont {Hollenberg}, \citenamefont {Du}, \citenamefont {Liu},\ and\
  \citenamefont {Wrachtrup}}]{Reinhard2012}%
  \BibitemOpen
  \bibfield  {author} {\bibinfo {author} {\bibfnamefont {F.}~\bibnamefont
  {Reinhard}}, \bibinfo {author} {\bibfnamefont {F.}~\bibnamefont {Shi}},
  \bibinfo {author} {\bibfnamefont {N.}~\bibnamefont {Zhao}}, \bibinfo {author}
  {\bibfnamefont {F.}~\bibnamefont {Rempp}}, \bibinfo {author} {\bibfnamefont
  {B.}~\bibnamefont {Naydenov}}, \bibinfo {author} {\bibfnamefont
  {J.}~\bibnamefont {Meijer}}, \bibinfo {author} {\bibfnamefont {L.~T.}\
  \bibnamefont {Hall}}, \bibinfo {author} {\bibfnamefont {L.}~\bibnamefont
  {Hollenberg}}, \bibinfo {author} {\bibfnamefont {J.}~\bibnamefont {Du}},
  \bibinfo {author} {\bibfnamefont {R.-B.}\ \bibnamefont {Liu}}, \ and\
  \bibinfo {author} {\bibfnamefont {J.}~\bibnamefont {Wrachtrup}},\ }\href
  {\doibase 10.1103/PhysRevLett.108.200402} {\bibfield  {journal} {\bibinfo
  {journal} {Phys. Rev. Lett.}\ }\textbf {\bibinfo {volume} {108}},\ \bibinfo
  {pages} {200402} (\bibinfo {year} {2012})}\BibitemShut {NoStop}%
\bibitem [{\citenamefont {Zhao}\ \emph {et~al.}(2011)\citenamefont {Zhao},
  \citenamefont {Wang},\ and\ \citenamefont {Liu}}]{Zhao2011}%
  \BibitemOpen
  \bibfield  {author} {\bibinfo {author} {\bibfnamefont {N.}~\bibnamefont
  {Zhao}}, \bibinfo {author} {\bibfnamefont {Z.-Y.}\ \bibnamefont {Wang}}, \
  and\ \bibinfo {author} {\bibfnamefont {R.-B.}\ \bibnamefont {Liu}},\ }\href
  {\doibase 10.1103/PhysRevLett.106.217205} {\bibfield  {journal} {\bibinfo
  {journal} {Phys. Rev. Lett.}\ }\textbf {\bibinfo {volume} {106}},\ \bibinfo
  {pages} {217205} (\bibinfo {year} {2011})}\BibitemShut {NoStop}%
\bibitem [{\citenamefont {Huang}\ \emph {et~al.}(2011)\citenamefont {Huang},
  \citenamefont {Kong}, \citenamefont {Zhao}, \citenamefont {Shi},
  \citenamefont {Wang}, \citenamefont {Rong}, \citenamefont {Liu},\ and\
  \citenamefont {Du}}]{Huang2011}%
  \BibitemOpen
  \bibfield  {author} {\bibinfo {author} {\bibfnamefont {P.}~\bibnamefont
  {Huang}}, \bibinfo {author} {\bibfnamefont {X.}~\bibnamefont {Kong}},
  \bibinfo {author} {\bibfnamefont {N.}~\bibnamefont {Zhao}}, \bibinfo {author}
  {\bibfnamefont {F.}~\bibnamefont {Shi}}, \bibinfo {author} {\bibfnamefont
  {P.}~\bibnamefont {Wang}}, \bibinfo {author} {\bibfnamefont {X.}~\bibnamefont
  {Rong}}, \bibinfo {author} {\bibfnamefont {R.-B.}\ \bibnamefont {Liu}}, \
  and\ \bibinfo {author} {\bibfnamefont {J.}~\bibnamefont {Du}},\ }\href
  {\doibase 10.1038/ncomms1579} {\bibfield  {journal} {\bibinfo  {journal}
  {Nature Communications}\ }\textbf {\bibinfo {volume} {2}},\ \bibinfo {pages}
  {570} (\bibinfo {year} {2011})}\BibitemShut {NoStop}%
\bibitem [{\citenamefont {Clerk}\ \emph {et~al.}(2010)\citenamefont {Clerk},
  \citenamefont {Devoret}, \citenamefont {Girvin}, \citenamefont {Marquardt},\
  and\ \citenamefont {Schoelkopf}}]{Clerk2010}%
  \BibitemOpen
  \bibfield  {author} {\bibinfo {author} {\bibfnamefont {A.~A.}\ \bibnamefont
  {Clerk}}, \bibinfo {author} {\bibfnamefont {M.~H.}\ \bibnamefont {Devoret}},
  \bibinfo {author} {\bibfnamefont {S.~M.}\ \bibnamefont {Girvin}}, \bibinfo
  {author} {\bibfnamefont {F.}~\bibnamefont {Marquardt}}, \ and\ \bibinfo
  {author} {\bibfnamefont {R.~J.}\ \bibnamefont {Schoelkopf}},\ }\href
  {\doibase 10.1103/RevModPhys.82.1155} {\bibfield  {journal} {\bibinfo
  {journal} {Rev. Mod. Phys.}\ }\textbf {\bibinfo {volume} {82}},\ \bibinfo
  {pages} {1155} (\bibinfo {year} {2010})}\BibitemShut {NoStop}%
\bibitem [{\citenamefont {Fedorov}\ \emph {et~al.}(2011)\citenamefont
  {Fedorov}, \citenamefont {Macha}, \citenamefont {Feofanov}, \citenamefont
  {Harmans},\ and\ \citenamefont {Mooij}}]{Fedorov2011}%
  \BibitemOpen
  \bibfield  {author} {\bibinfo {author} {\bibfnamefont {A.}~\bibnamefont
  {Fedorov}}, \bibinfo {author} {\bibfnamefont {P.}~\bibnamefont {Macha}},
  \bibinfo {author} {\bibfnamefont {A.~K.}\ \bibnamefont {Feofanov}}, \bibinfo
  {author} {\bibfnamefont {C.~J. P.~M.}\ \bibnamefont {Harmans}}, \ and\
  \bibinfo {author} {\bibfnamefont {J.~E.}\ \bibnamefont {Mooij}},\ }\href
  {\doibase 10.1103/PhysRevLett.106.170404} {\bibfield  {journal} {\bibinfo
  {journal} {Phys. Rev. Lett.}\ }\textbf {\bibinfo {volume} {106}},\ \bibinfo
  {pages} {170404} (\bibinfo {year} {2011})}\BibitemShut {NoStop}%
\bibitem [{\citenamefont {Fink}\ and\ \citenamefont {Bluhm}(2013)}]{Fink2013}%
  \BibitemOpen
  \bibfield  {author} {\bibinfo {author} {\bibfnamefont {T.}~\bibnamefont
  {Fink}}\ and\ \bibinfo {author} {\bibfnamefont {H.}~\bibnamefont {Bluhm}},\
  }\href {\doibase 10.1103/PhysRevLett.110.010403} {\bibfield  {journal}
  {\bibinfo  {journal} {Phys. Rev. Lett.}\ }\textbf {\bibinfo {volume} {110}},\
  \bibinfo {pages} {010403} (\bibinfo {year} {2013})}\BibitemShut {NoStop}%
\bibitem [{\citenamefont {Barthel}\ \emph {et~al.}(2010)\citenamefont
  {Barthel}, \citenamefont {Kj\ae{}rgaard}, \citenamefont {Medford},
  \citenamefont {Stopa}, \citenamefont {Marcus}, \citenamefont {Hanson},\ and\
  \citenamefont {Gossard}}]{Barthel2010}%
  \BibitemOpen
  \bibfield  {author} {\bibinfo {author} {\bibfnamefont {C.}~\bibnamefont
  {Barthel}}, \bibinfo {author} {\bibfnamefont {M.}~\bibnamefont
  {Kj\ae{}rgaard}}, \bibinfo {author} {\bibfnamefont {J.}~\bibnamefont
  {Medford}}, \bibinfo {author} {\bibfnamefont {M.}~\bibnamefont {Stopa}},
  \bibinfo {author} {\bibfnamefont {C.~M.}\ \bibnamefont {Marcus}}, \bibinfo
  {author} {\bibfnamefont {M.~P.}\ \bibnamefont {Hanson}}, \ and\ \bibinfo
  {author} {\bibfnamefont {A.~C.}\ \bibnamefont {Gossard}},\ }\href {\doibase
  10.1103/PhysRevB.81.161308} {\bibfield  {journal} {\bibinfo  {journal} {Phys.
  Rev. B}\ }\textbf {\bibinfo {volume} {81}},\ \bibinfo {pages} {161308}
  (\bibinfo {year} {2010})}\BibitemShut {NoStop}%
\bibitem [{\citenamefont {Cywi\ifmmode~\acute{n}\else \'{n}\fi{}ski}\ \emph
  {et~al.}(2009)\citenamefont {Cywi\ifmmode~\acute{n}\else \'{n}\fi{}ski},
  \citenamefont {Witzel},\ and\ \citenamefont {Das~Sarma}}]{Cywinski2009}%
  \BibitemOpen
  \bibfield  {author} {\bibinfo {author} {\bibfnamefont {L.}~\bibnamefont
  {Cywi\ifmmode~\acute{n}\else \'{n}\fi{}ski}}, \bibinfo {author}
  {\bibfnamefont {W.~M.}\ \bibnamefont {Witzel}}, \ and\ \bibinfo {author}
  {\bibfnamefont {S.}~\bibnamefont {Das~Sarma}},\ }\href {\doibase
  10.1103/PhysRevB.79.245314} {\bibfield  {journal} {\bibinfo  {journal} {Phys.
  Rev. B}\ }\textbf {\bibinfo {volume} {79}},\ \bibinfo {pages} {245314}
  (\bibinfo {year} {2009})}\BibitemShut {NoStop}%
\bibitem [{Note1()}]{Note1}%
  \BibitemOpen
  \bibinfo {note} {See Supplemental Material at [URL] for a detailed derivation
  of the autocorrelation function for the switched unixaxial coupling and
  isotropic coupling with dynamically decoupling pulses scheme.}\BibitemShut
  {Stop}%
\end{thebibliography}%

\newpage
\onecolumngrid
\appendix
The following supplemental material is divided into two sections which provide the derivation of the autocorrelation function for the switched uniaxial coupling, and the isotropic coupling with dynamical decoupling pulses, respectively.

\section{Switched uniaxial coupling}
We start the derivation of the autocorrelation function for the singlet-triplet qubit from Eq.~(5) of the main text and first consider only one electron spin coupled to a nuclear spin bath. The noncommuting evolution operators $\hat{U}_{0,1} = \otimes_i^N \exp \left( \mp \frac{1}{2} A_i \hat{J}_i^z t_M \right)$ and $\hat{U}_I = \otimes_i^N \exp \left( - \frac{1}{2} A_i \hat{J}_i^x t_I \right)$ factorize into single spin terms and can thus be evaluated analytically. To account for inhomogeneous coupling between the electron and nuclear spins, we assume an electron wave function of the form
\begin{equation}
\psi(\mathbf{r})=\sqrt{\frac{2\nu_0}{z_0 \pi L^2}}\cos\left(\frac{\pi z}{z_0}\right)\exp\left(-\frac{x^2+y^2}{2L^2}\right),
\end{equation}
where $z_0$ is the dot thickness, $\nu_0$ the volume of the primitive cell, and $L$ the Fock-Darwin radius, which is a suitable description for electrically confined electrons in a quantum well. Similar to Ref.\,\cite{Cywinski2009}, we then group the nuclear spins into $n$ clusters where the coupling of all nuclei within a single cluster is given by the nuclear spin-specific constant $\mathcal{A}$ weighted by the electron wave function. This approach allows efficient averaging over all bath states. We continue to neglect the internal bath dynamics, which can be justified in retrospect by comparison to experimental coherence times \cite{Cywinski2009}. To calculate expectation values of the autocorrelation function, we average over initial nuclear spin bath configurations where the bath polarization follows a binomial distribution with probability $\frac{1}{2}$ corresponding to an unpolarized bath. Note, however, that also polarized baths can be implemented by adjusting the probability according to the ratio of the nuclear spin states. When the baths of the two dots are independent, the two terms of the autocorrelation function in Eq.~(5) of the main text can be decomposed into products where each factor contains operators acting only on one dot. Moreover, we assume that both dots contain an equal number of nuclear spins and that the electron wave functions are identical, i.e. symmetric dots. In this case, the contributions from the two dots are equal and only lead to a square of the autocorrelation function as compared to the single electron case. The autocorrelation function then reads
\begin{eqnarray} \nonumber \label{eq:corrtoy}
\langle C \rangle &=& \frac{1}{2} \prod_{i=1}^n \left| \sum_{k=1}^{N_i} \frac{1}{2^{N_i}} \binom{N_i}{k} \left[ \langle \uparrow | \hat{U}_i^\prime | \uparrow \rangle \right]^k \left[ \langle \downarrow | \hat{U}_i^\prime | \downarrow \rangle \right]^{N_i-k} \right|^2 \\
&+& \frac{1}{2} \prod_{i=1}^n \left| \sum_{k=1}^{N_i} \frac{1}{2^{N_i}} \binom{N_i}{k} \left[ \langle \uparrow | \hat{U}_i^{\prime \prime} | \uparrow \rangle \right]^k \left[ \langle \downarrow | \hat{U}_i^{\prime \prime} | \downarrow \rangle \right]^{N_i-k} \right|^2,
\end{eqnarray}
where the sum over $i$ and $k$ runs over the hyperfine coupling clusters and nuclear spins, respectively, and $\hat{U}_i^\prime$ ($\hat{U}_i^{\prime \prime}$) are the single-spin operators of the sequence of operators in the first (second) term of Eq.~(5) of the main text, effectively rotating each nuclear spin about the $z$- and $x$-axis. These terms are given by
\begin{eqnarray} \label{eq:spinterms} \nonumber
\langle \uparrow_i | \hat{U}_i^\prime |\uparrow_i \rangle &=& \exp \left( -i\frac{A_i}{2}t_M \right) \left[-2i\sin \left( \frac{A_i}{2}t_M \right) \cos^2 \left( \frac{A_i}{4}t_I \right) + \exp \left( i \frac{A_i}{2}t_M \right) \right] \\
\langle \uparrow_i | \hat{U}_i^{\prime \prime} |\uparrow_i \rangle &=& \exp \left( i\frac{A_i}{2}t_M \right) \left[-2i\sin \left( \frac{A_i}{2}t_M \right) \cos^2 \left( \frac{A_i}{4}t_I \right) + \exp \left( i \frac{A_i}{2}t_M \right) \right].
\end{eqnarray}
It can further be shown that the spin up and spin down-terms in Eq.~\ref{eq:spinterms} are complex conjugates
\begin{equation}
\langle \uparrow_i | \hat{U}_i^{\prime,\prime \prime} | \uparrow_i \rangle = \left( \langle \downarrow | \hat{U}_i^{\prime,\prime \prime} | \downarrow \rangle \right)^*.
\end{equation}
Evaluation of Eq.~\ref{eq:corrtoy} then allows to calculate expectation values of the autocorrelation function for this scheme. The result is shown in Fig.~2 of the main text for inhomogeneous coupling of both electron spins to nuclear spin baths each consisting of $10^3$ nuclei of a single species. Other isotopes can easily be included into the calculation by adjusting the coupling strength for those nuclei.

\section{Isotropic coupling with dynamically decoupling pulses}
For the deviation of the autocorrelation function, we will treat the nuclei as classical vectors with electron spin-dependent dynamics. To calculate the expectation value of the correlation function, we will average over initial nuclear spin configurations \cite{Neder2011}. We again assume that the baths are independent, which allows us to decompose the correlation function into factors describing only one dot. We therefore start by considering a single electron spin coupled to a bath. In this case, the autocorrelation is determined via the phases $\phi_1$ and $\phi_2$ accumulated during the first and second SE measurement, respectively. The autocorrelation is then given by
\begin{equation} \label{eq:correlation}
\langle C \rangle = \frac{1}{2} \Re \left( \left\langle e^{i(\phi_1 + \phi_2)} \right\rangle + \left\langle e^{i(\phi_1 - \phi_2)} \right\rangle \right).
\end{equation}
These phases measure the dynamics of the Overhauser field created by the nuclear spins and experienced by the electron. For a SE measurement, the longitudinal field component $B_{nuc}^z$, which is assumed to be constant for typical qubit evolution times, cancels out and the measurement is only sensitive to the transverse Overhauser field components
\begin{equation}
|\mathbf{B}^\bot_{nuc}(t)|^2 = \sum_{k,l}b_k(t)b_l^* (t),
\end{equation}
where the sum runs over the different nuclear field components, such as different Larmor frequencies $\omega_k$ or hyperfine couplings $A_k$. The dynamics of the complex Overhauser field components $b_k = b_k^x + i b_k^y$ during one evolution period, possibly under the influence of dynamically decoupling pulses, can be expressed as
\begin{equation} \label{eq:overhauser}
b_k^\sigma (t) = b_k^\sigma(0) \exp \left[i \left( (\omega_k + \delta \omega_k) t + \sigma A_k \int_0^{t} dt^\prime c(t^\prime) \right) \right],
\end{equation}
where we have introduced the index $\sigma=\pm \frac{1}{2}$ to denote the electron spin state, and $b_k^\sigma	(0)$ is the Overhauser field component at the beginning of the respective evolution. The term $\delta \omega_k$ phenomenologically accounts for nuclear spin dephasing via normally distributed local magnetic field fluctuations $\delta B_{ext}$ \cite{Neder2011}. The function $c(t)=\pm 1$ switches its sign whenever a $\pi$-pulse is applied and thus captures the effect of any dynamical decoupling pulses applied. Note how a switching at $t/2$ cancels out any electron spin-dependent dynamics (i.e. backaction) of the Overhauser field components.

The phases in Eq.~\ref{eq:correlation} can now be calculated via
\begin{equation} \label{eq:phase}
\phi_{1,2} = C \sum_{k,l} b_k^\sigma(0) b_l^{\sigma*}(0) \sum_{\sigma = \pm \frac{1}{2}} \int_0^{t_M} dt^\prime c(t^\prime) e^{i\omega_{kl}t^\prime + i\sigma A_{kl} \int dt^{\prime\prime} c(t^{\prime\prime})},
\end{equation}
where $C=\frac{g\mu_B}{4|B_{ext}|}$ is a constant determined by the electron gyromagnetic factor and external magnetic field, and $\omega_{kl}=\omega_k - \omega_l$ ($A_{kl} = A_k - A_l$) denotes the relative Larmor frequencies (hyperfine couplings). Note that for $\phi_2$, the initial Overhauser field components $b_{k,l}^\sigma (0)$ incorporate the potential backaction due to the intermediate FID/SE as described in Eq.~\ref{eq:overhauser}. The intermediate FID thus imprints an additional phase onto the transverse Overhauser field components, effectively defining the initial Overhauser field configuration for the second measurement. Due to inhomogeneous coupling, this imprinted phase then leads to a suppression of the peaks arising from the relative Larmor precessions in Eq.~\ref{eq:phase}. When the initial Overhauser fields are expressed via their rms value $\bar{b}_k = \frac{\hbar A_k}{2g \mu_B} \sqrt{5 N_k}$, where $N_k$ is the number of nuclei of species $k$ in the dot, and complex, normally distributed variables $z_k$, the phases may be rewritten via a $T$-matrix using the identity $\left\langle e^{i(\phi_1^\sigma \pm \phi_2^\sigma)} \right\rangle = \left\langle e^{-\frac{i}{2} \sum_{k,l} T_{k,l}^{\pm,\sigma} z_k z_l^*} \right\rangle$, where
\begin{equation} \label{eq:tmatrix}
T_{kl}^{\pm,\sigma} = \frac{4ig\mu_B\bar{b}_k \bar{b}_l}{\hbar B_{ext}} \frac{\exp \left( i\frac{\omega_{kl}+\delta \omega_{kl}}{2}t_M\right)}{i(\omega_{kl}+\delta \omega_{kl})} \sin^2 \left( \frac{\omega_{kl}+\delta \omega_{kl}}{4} t_M \right) \left( 1\pm e^{i((\omega_{kl}+\delta \omega_{kl})(t_M+t_I) + \sigma A_{kl}t_I)} \right).
\end{equation}
We now take into account the double dot-configuration which simply leads to products of terms where each factor acts on only one dot. Performing the average in Eq.~\ref{eq:correlation} by averaging over the normally distributed complex $z_{k,l}$, we derive the autocorrelation function
\begin{eqnarray} \nonumber
\langle C \rangle &=& \frac{1}{4} \Re \left[ \prod_m \frac{1}{1+i\lambda_{m,L}^{+,\uparrow}} \frac{1}{1-i\lambda_{m,R}^{+,\downarrow}} + \prod_m \frac{1}{1+i\lambda_{m,L}^{+,\downarrow}} \frac{1}{1-i\lambda_{m,R}^{+,\uparrow}} \right] \\
&+& \frac{1}{4} \Re \left[ \prod_m \frac{1}{1+i\lambda_{m,L}^{-,\uparrow}} \frac{1}{1-i\lambda_{m,R}^{-,\downarrow}} + \prod_m \frac{1}{1+i\lambda_{m,L}^{-,\downarrow}} \frac{1}{1-i\lambda_{m,R}^{-,\uparrow}} \right],
\end{eqnarray}
where $L,R$ denote the left and right dot, and $\lambda_{m,d}^{\pm,\sigma}$ are the $m$ eigenvalues of the T-matrix defined in Eq.~\ref{eq:tmatrix} \cite{Neder2011}. For symmetric dots, this expression can again be simplified to
\begin{equation}
\langle C \rangle = \frac{1}{2} \Re \sum_{s=\pm} \prod_m \frac{1}{1+i\lambda_{m}^{s,\uparrow}} \frac{1}{1-i\lambda_{m}^{s,\downarrow}}.
\end{equation}
The result of this function for a singlet-triplet qubit with a spin bath consisting of $10^6$ nuclei is illustrated in Fig.~3 of the main text.

\end{document}